\documentclass[reprint,secnumarabic, amssymb, amsmath, aps,prl]{revtex4-1}
\usepackage{graphicx,bm}
\usepackage[colorlinks=true,linkcolor=blue]{hyperref}%
\expandafter\ifx\csname package@font\endcsname\relax\else
\expandafter\expandafter
\expandafter\usepackage
\expandafter\expandafter
\expandafter{\csname package@font\endcsname}%
\fi
\begin{document}

\title{Comment on ''Negative Landau damping in bilayer graphene''}

\author{Dmitry Svintsov}%
\affiliation{Laboratory of 2D Materials' Optoelectronics, Moscow Institute of Physics and Technology, Dolgoprudny 141700,	Russia}%

\author{Victor Ryzhii}%
\affiliation{Research Institute of Electrical Communication, Tohoku University, Sendai 980-8577\,Japan}%

\maketitle

In Ref.~\onlinecite{Silveirinha} it was argued that two parallel graphene layers in the presence of electron drift support unstable plasmon modes. Here we show that the predicted instability is an artifact of two errors in the calculation of graphene polarizability $\Pi({\bf q},\omega)$ (here $\bf q$ is the wave vector and $\omega$ is the frequency) (1) long-wavelength expansion of $\Pi$ in the short-wavelength domain $\omega \lesssim q v_0$ ($v_0$ is the electron Fermi velocity) (2) application of Doppler shift $\omega \rightarrow \omega - {\bf q u}_0$ for the evaluation of polarizability in the presence of drift with velocity ${\bf u}_0$ (Eq.~(1) of [\onlinecite{Silveirinha}]). The latter is invalid in graphene~\cite{VanDuppen_birefringent,Hydrodynamic_transport} due to the absence of Galilean invariance~\cite{Abedinpour,Levitov_plasmons}.

Upon renouncement of long-wavelength expansion, graphene polarizability acquires a square-root singularity at the threshold of single-particle excitations (SPEs) ${\rm Re}\Pi({\bf q},\omega)\propto [\omega^2 - q^2 v_0^2]^{-1/2}$~\cite{Das_Sarma_Plasmons,Ryzhii-plasmons}. This singularity keeps plasmon phase velocity above Fermi velocity, $\omega/q > v_0$ (see recent experiment \cite{Lundeberg_Nonlocal_Plasmons}), and protects the waves from Landau damping. Below we show that such singularity persists at finite drift velocity. Therefore, the frequency of graphene plasmons cannot be pushed into the domain of negative Landau damping $\omega < q u_0$ by current, contrary to the case of massive two-dimensional systems~\cite{krasheninnikov1980instabilities}.

The proper calculation of $\Pi(q,\omega)$ exploits Lindhard (RPA) formula with drifting electron distribution function $f_{\bf p} = [1+e^{(pv_0 - {\bf p u}_0-\varepsilon_F)/T}]^{-1}$. We set $\hbar = k_B = 1$ and restrict ourselves to the classical limit ($\omega\ll \varepsilon_F$) where the polarizability reads:
\begin{equation}
\label{Lindhard}
\Pi ({\bf q},\omega)= 4 \int{\frac{d^2{\bf p}}{(2\pi)^2}}{\frac{{\bf q} \nabla_{\bf p} f_{{\bf p}} }{ \omega + i \delta - {\bf q v}_{\bf p}}};
\end{equation}
here ${\bf v}_{\bf p}=v_0{\bf p}/p$ is the carrier velocity. The integral is evaluated analytically using the residue theorem~\cite{Kukhtaruk}. We present the final result for waves propagating parallel ($\Pi_+$) and anti-parallel ($\Pi_-$) to the drift direction:
\begin{equation}
\label{P-drift}
    \Pi_\pm = \frac{2}{\pi}\frac{\varepsilon_F}{(1\mp s\beta)^2}\left(\sqrt{1-\beta^2} - \frac{s\pm \beta}{\sqrt{s^2-1}} \right);
\end{equation}
we have denoted $s=(\omega+i\delta)/qv_0$ and $\beta = u_0/v_0$~\footnote{At finite temperature, Fermi energy should be replaced according to $\varepsilon_F \rightarrow T \ln(1+e^{\varepsilon_F/T})$}. This expression is radically different from Doppler-shifted polarizability in the $q\rightarrow 0$ limit used in~[\onlinecite{Silveirinha}], 
\begin{equation}
\label{P-approx}
    \tilde \Pi_\pm \approx -\frac{\varepsilon_F}{\pi}\left(\frac{qv_0}{\omega\pm qu_0}\right)^2
\end{equation}
The persistence of square-root singularity at $\omega = qv_0$ and the lack of Galilean invariance are apparent from Eq.~(\ref{P-drift}). The singularity persists in full-quantum Lindhard formalism~\cite{VanDuppen_birefringent} and upon different assumptions about the form of distribution function (e.g. shifted Fermi disc model~\cite{Sabbaghi_Drift-induced}).

We substitute the polarization (\ref{P-drift}) into the dispersion relation for plasmons in double graphene layer (Eq. (2) of [\onlinecite{Silveirinha}]) and observe that it has no unstable roots (Fig. 1), in contrast to the dispersion with approximated polarization~(\ref{P-approx}). The proper plasmon frequencies retain outside of SPE domain at finite $u_0$ and thus have zero imaginary part. Inclusion of interband polarizability and/or electron collisions makes these modes decaying. We have verified that instabilities are absent at all accessible drift velocities $u_0<v_0$, Fermi energies, and propagation angles. The instabilities might re-appear in hydrodynamic regime where the singularity at $\omega=qv_0$ is removed~\cite{Hydrodynamic_transport}. However, calculations in this regime require the methods essentially different from used both in Ref.~\onlinecite{Silveirinha} and here.

\begin{figure}[ht]
	\includegraphics[width=1.0\linewidth]{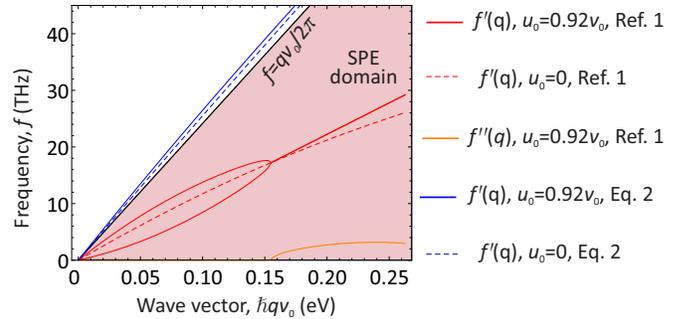}
	\caption{\label{Fig1} 
		Plasmon dispersions for double graphene layer obtained with the formalism of Ref.~\cite{Silveirinha} (Eq.~(\ref{P-approx}), red and orange lines) and with correct polarization, Eq.~\ref{P-drift} (blue line). Dashed lines correspond $u_0=0$, solid lines to $u_0=0.92v_0$. Filled area is the SPE continuum. Parameters: $\varepsilon_F = 0.1$ eV, interlayer gap $d=5$ nm, gap and substrate dielectric constants $\epsilon_g=12$ and  $\epsilon_s=4$}
\end{figure}


In conclusion, the plasmon instability predicted in [\onlinecite{Silveirinha}] is an artifact of unjustified approximations to graphene polarizability. Accurate treatment shows that the collisionless electron plasma in graphene double layers is stable at arbitrary drift velocity.

This work was supported by Grant No. 16-19-10557 of the Russian Science Foundation.

\bibliography{Petrov_Autumn_2016}

\end{document}